\documentclass[pra,showpacs]{revtex4}
\usepackage{graphicx}
\usepackage{bm}
\begin{document}
\title{Scaling of entanglement at quantum phase transition for two-dimensional array of quantum dots}
\author{Jiaxiang Wang and Sabre Kais}
\affiliation{Chemistry Department, Purdue University, West Lafayette, IN 47907}

\begin{abstract}
With Hubbard model, the entanglement scaling behavior in a two-dimensional itinerant system is 
investigated. It has been found that, on the two sides of the critical point denoting an inherent
quantum phase transition (QPT), the entanglement follows different scalings with the size just as an order
parameter does. This fact reveals the subtle role played by the entanglement in QPT as a fungible physical resource. 
\end{abstract}
\maketitle

The existence of entanglement between distinctive quantum systems has marked a fundamental difference between quantum and classical physics. Recently, with the explosive development of the research in quantum information theory and quantum computation \cite{vandersypen,bouwmeester,jennewein,Mattle}, the study of the entanglement \cite{Lewenstein,bruss} has come to the limelight again after more than 60 years of controversies and strenuous progress. Experimentally,  entanglements have already been
produced between  up to four photons \cite{lamas,eibl} and even between two
macroscopic states such as two superconducting qubits, each of which contains as large as $10^{9}$
electrons \cite{berkley}. But theoretically, because an ensemble's Hilbert space
grows exponentially with the number of its component particles, we are still far
from fully understanding the contents of the entanglements. Only for the simplest state
with two distinguishable particles can we have a complete description of the
entanglement measure. For states of more than two particles, especially for
mixed states, the current knowledge about their entanglement is very
limited and all the related complexities have just begun to be explored. For spin-only entanglement of localized distinguishable particles, the most popular measure of the entanglement is the Wootters'
measure \cite{wootters}. Recently, the influence of the quantum statistics on the definition of the entanglement has
begun to be noticed and discussed by several authors \cite{schliemann,li,ahronov}.    
Although various entanglement measures have been put forward, according to Gitting's criterions
\cite{gittings}, only Zanardi's measure \cite{zanardi} survives the test of all the requirements upon entanglement
definition. This measure is given in Fock space as the von Neuman entropy, namely,

\begin{equation}
 E_j =  - Tr\rho _j \ln \rho _j, \;\;\;\; \rho _j  =Tr_j \left| \psi  \right\rangle  \left\langle \psi  \right|,
\label{entanglement} 
\end{equation}
where $Tr_j$ denotes the trace over all but the $jth$ site and $\psi$ is the antisymmetric
wave function of the studied system. Hence $E_j$ actually describes the
entanglement of the $jth$ site with the remaining sites. 
A generalization of this one-site  entanglement is to define an entanglement between one
L-site block with the rest of the systems \cite{vidal},

\begin{equation}
E_L  =  - Tr(\rho _L \log _2 \rho _L ).
\label{entanglement1}
\end{equation}
where all the sites are traced out except those belonging to the selected block.

Recently, it has been speculated that the most entangled systems could be found at 
the critical point when the system undergoes a quantum phase transition (QPT), i.
e. a qualitative change of some physical properties takes place as an order parameter in the Hamiltonian is tuned
\cite{sachdev}. QPT results from quantum fluctuations at the absolute zero of temperature and is a pure quantum
effect featured by long-range correlations. So far, there have already been some efforts in exploring the above
speculations, such as the analysis of the  XY model about the single-spin entropies and two-spin quantum correlations \cite{osborne1,kais}, the entanglement between a block of $L$ contiguous sites and the rest of the chain \cite{vidal} and also the scaling of entanglement near QPT \cite{osterloh}. But because there is still no
analytical proof, the role played by the entanglement in quantum critical phenomena remains elusive. Generally speaking, there exists  at least two difficulties in resolving this issue. First, until now, only two-particle entanglement is well explored. How to quantify the multi-particle entanglement is not clear.  Second, QPT closely relates to the notorious many-body problems, which is almost intractable analytically. Until now, the only effective
and accurate way to deal with QPT in critical region is the density-matrix renormalization group method
\cite{white}. Unfortunately, it is only efficient for one-dimensional cases because of the much more complicated boundary conditions for two-dimensional situation. It should be mentioned here that recently, Vial \cite{vidal1} has put forward another new efficient numerical method to study one-dimensional many-body systems based upon the entanglement contained in the system.

In this paper, we will focus on investigating the entanglement behavior in QPT for two-dimensional array of
quantum dots, which provide a suitable arena for implementation of quantum
computation\cite{loss1,losss2,sarma}. For this purpose, the real-space renormalization group
technique\cite{realspace} will be utilized and developed for the finite-size analysis of entanglement. 

The model we use is the Hubbard model with the Hamiltonian,
\begin{eqnarray}
H=&&-t\sum_{<i,j>,\sigma }[c_{i\sigma }^{+}c_{j\sigma}+H.c.]\nonumber \\
&&+U\sum_{i}(\frac{1}{2}-n_{i\uparrow })(\frac{1}{2}-n_{i\downarrow })+K\sum_{i}I_{i} 
\end{eqnarray}
where $t$ is the nearest-neighbor hopping term, $U$ is the local repulsive interaction and $K=-U/4$ and $I_i$ is the 
unit operator. $c_{i\sigma}^{+}(c_{i\sigma })$ creates(annihilates) an electron with spin $\sigma $ in a Wannier orbital located at site $i$; the corresponding number operator is $n_{i\sigma }=c_{i\sigma }^{+}c_{i\sigma }$ and 
$<>$ denotes the nearest-neighbor pairs. H.c. denotes the Hermitian conjugate.

For a half-filled triangular quantum lattice, there exists a metal-insulator phase transition with the tuning
parameter $U/t$ at the critical point $12.5$ \cite{wang1,wang2,wang3}. The
corresponding order parameter for metal-insulator transition is the charge gap defined by $\triangle _{g}=E(N_{e}-1)+E(N_{e}+1)-2E(N_{e})$, where $E(N_{e})$ denotes the lowest energy for a $N_{e}-$electron system. In 
our case, $N_{e}$ is equal to the site number $N_{s}$ of the lattice. 
Unlike the charge gap calculated from the energy levels, the Zanardi's measure of the entanglement
is defined upon the wave function corresponding to $E(N_e)$ instead. Using the conventional renormalization group
method for the finite-size scaling analysis\cite{wang1,wang2,wang3}, we can discuss three schemes of entanglement scaling:

1)Single-site entanglement scaling with the total system size,$E_{single}$;

2)Single-block entanglement scaling with the block size, $E_{block}$; and 

3)Block-block entanglement scaling with the block size, $E_{block-block}$.

Fig. 1 presents the single-site entanglement scaling. It is obvious that $E_{single}$ is not a universal quantity. 
This conclusion is well consistent with the argument given by Osborne \cite{osborne1}, who claims that the 
single-site entanglement is not scalable because it does not own the proper extensivity and does not distinguish the 
local and distributed entanglement. One more interesting feature in Fig. 1 is that when the  system size is increased beyond $7^2$, $E_{single}$ almost makes no change any
more. 

This implies that only a limited region of sites around the central site contributed significantly to
the single-site entanglement. Using the one-parameter scaling theory, near the phase transition
point, we assume the existence of scaling function $f$ for $E_{block-block}$ such that:
\begin{eqnarray}
E_{block-block} =q^{y_E} f(\frac{L}{\xi })
\end{eqnarray}
where $q=U/t-(U/t)_c$ measures the deviation distance of the system away from the critical state with
$(U/t)_c=12.5$, which is exactly equal to the critical value for metal-insulator transition  when the same order parameter $U/t$ is used \cite{wang1,wang2,wang3}. $\xi  = q^{ - \nu }$  is the correlation length of the system
with  the critical exponent $\nu$. 

Hence,
\begin{eqnarray}
E_{block-block} = q^{y_E } f(N^{\frac{1}{{2\nu}}} q),
\end{eqnarray} 
where we used  $N=L^2$ for the two-dimensional systems.

In Fig. 2, we show the results of $E_{block-block}$ as a function  of $(U/t)$ for different system sizes. 
With proper scaling, all the curves collapse onto one curve, which can be expressed as
\begin{eqnarray}
E_{block-block} = f(qN^{\frac{1}{2}}).
\end{eqnarray}
Thus the critical exponents in Eq. (5) are  $y_E=0,\nu=1$. It is interesting to note that we obtained the same
$\nu$ as in the study of metal-insulator transition. This shows the consistency of the results since the critical exponent $\nu$ is only dependent on the inherent symmetry and dimension of the investigated system.

Another significant result lies in the finding that the metal state is highly entangled while the insulating state, 
is only partly entangled. For a $4$ dimensional density matrix, the maximally entangled state  can be written as a diagonal matrix with equal components $\frac 14$. The related entanglement is $- \sum\limits_{i = 1}^4 {\frac{1}{4}\log _2 \frac{1}{4}}  = 2$, which is exactly the value obtained from Fig.2(b). However, unlike the 
metal state, the insulating states should be expected to have electrons showing less mobility. If we assume the highly probable situation, i.e. the central site has equal probability to be in $\left| \uparrow  \right\rangle $, 
$\left| \downarrow  \right\rangle $ and no occupation in $\left|  0  \right\rangle $, $\left| \uparrow
\downarrow \right\rangle $, the corresponding entanglement is then $E_{block-block}=-\sum\limits_{i = 2}^3
{\frac{1}{2}\log _2 \frac{1}{2}}  = 1$, also consistent with the results from Fig. 2(b).

All the above discussions are confined to the entanglement between the central block and its surrounding blocks. 
Because the central block is a very special one showing the highest symmetry, one may wonder what can happen to the neighboring blocks, for example, the entanglement between block 7 and the rest 6 blocks. To answer this question, the same calculations are conducted and the results are the same except that in the metal state, the 
maximal entanglement is a little less than 2 and the minimal one is a little less than 1. This can be explained by the asymmetric position of site 1 in the block. 

It should be mentioned that the calculated entanglement here has a corresponding critical exponent $y_E=0$. This means that the entanglement is constant at the critical point over all sizes of the system. But it is not a constant over all values of $U/t$. There is an abrupt jump across the critical point as $L \to \infty$.

 If we divide the regime of the order parameter into non-critical regime
and critical regime, the results can be summarized as follows: In the non-critical regime, i.e. $U/t$ is away
from $(U/t)_c$, as $L$ increases, the entanglement will saturate  onto two different values depending on the 
sign of $U/t-(U/t)_c$; At the critical point, the entanglement is actually a constant independent of the size $L$. 

These properties are qualitatively different from the single-site entanglement discussed by Osborne \cite{osborne1}, where the entanglement with Zanardi's measure increases from zero
to the maximum at the critical point and then decreases again to zero as the order parameter $\gamma$ for $XY$
mode is tuned. 

These peculiar properties of the entanglement we have found here can be of potential interest to make an
effective ideal "entanglement switch". For example, with seven blocks of quantum dots on triangular lattice, the
entanglement among the blocks can be regulated as $"0"$ or $"1"$ almost immediately once the tuning parameter
$U/t$ crosses the critical point. The switch errors will depend on the size of the blocks. Since it has already
been a well-developed  technique to change $U/t$ for quantum dot lattice\cite{sarma,rafi}, the above scheme should
be workable. To remove the special confinement we have made upon the calculated entanglement, namely only the entanglement of block 1 and block 7 with the rest ones are considered, in the following, we will prove that the average pairwise entanglement also has the properties shown in Fig. 2(b).

Let $E_{block-block, 7}$ denotes the entanglement between the $7$th and all the remaining blocks in a hexagonal 
system. From the symmetry of the system we can show that the total pairwise entanglement $E_{tot}=6E_{block-block}  + 3E_{block-block, 7} $. The average 2-site entanglement is $E_{average}=E_{tot}/21 =(2E_{block-block}+E_{block-block,7} )/7$. Because 

\begin{eqnarray}
E_{block,7}  = g_1(qN^{\frac{1}{2}} ), \;\;\;\; 
E_{block,1}  = g_2(qN^{\frac{1}{2}} ), 
\end{eqnarray}
then we should have $E_{average}  = g(qN^{1/2})$, where $g_1,g_2$ and $g$ are scaling functions.

For obtaining the single-block entanglement, the first step is to make a cutoff over the system size. In our work, 
we let it  be $7^9$. The results are presented in Fig. 2(c). It is magnificent that as we change the size of the 
central block, its entanglement with all the rest sites follows the same scaling properties as $E_{block-block}$. It 
is understandable if we consider the fact that only a limited region round the block contribute mostly to 
$E_{block}$. This result greatly facilitate the fabrication of realistic entanglement control devices, such as 
quantum gates for quantum computer, since we don't need to delicately care about the number of component blocks in 
fear that the next neighboring or the next-next neighboring quantum dots should influence the switching effect.

In summary, in this paper, various schemes of the finite-size scaling properties with Zanardi's measure of entanglement has been investigated for Hubbard model on a triangular quantum lattice. The critical exponent $\nu=1$ has been found, which coincides well with our previous work in studying a quite different physical property, the charge gap. When the block size $L \to \infty$, the entanglement shows an abrupt change when the tuning parameter crosses the phase transition point. This property might be well applied to make an "entanglement switch" and shows the promising prospect of regarding entanglement as a new physical resource.

{\bf Acknowledgments}

We would like to acknowledge the financial support of
the National Science Foundation

\newpage
\begin{figure}[tbp]
\includegraphics{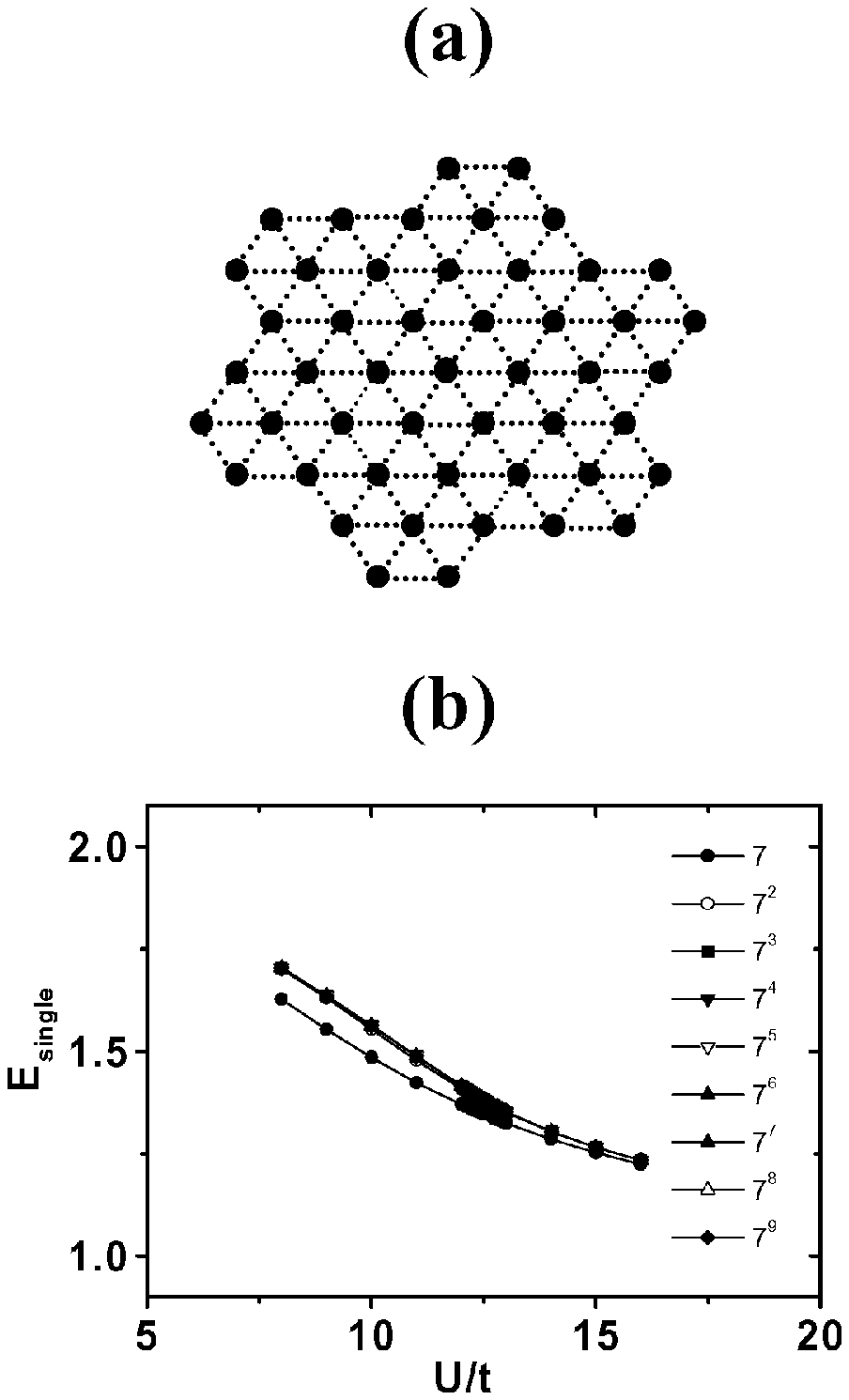}
\caption{Figure 1: (a) Schematic diagram showing the central site (purple) and the surrounding ones in the triangular quantum dot lattice. The dotted lines represent the site-site interactions. (b) Scaling of the single-site entanglement for various system size. The sizes are denoted by different symbols.}
\end{figure}

\newpage
\begin{figure}[tbp]
\includegraphics[width=0.80\textwidth]{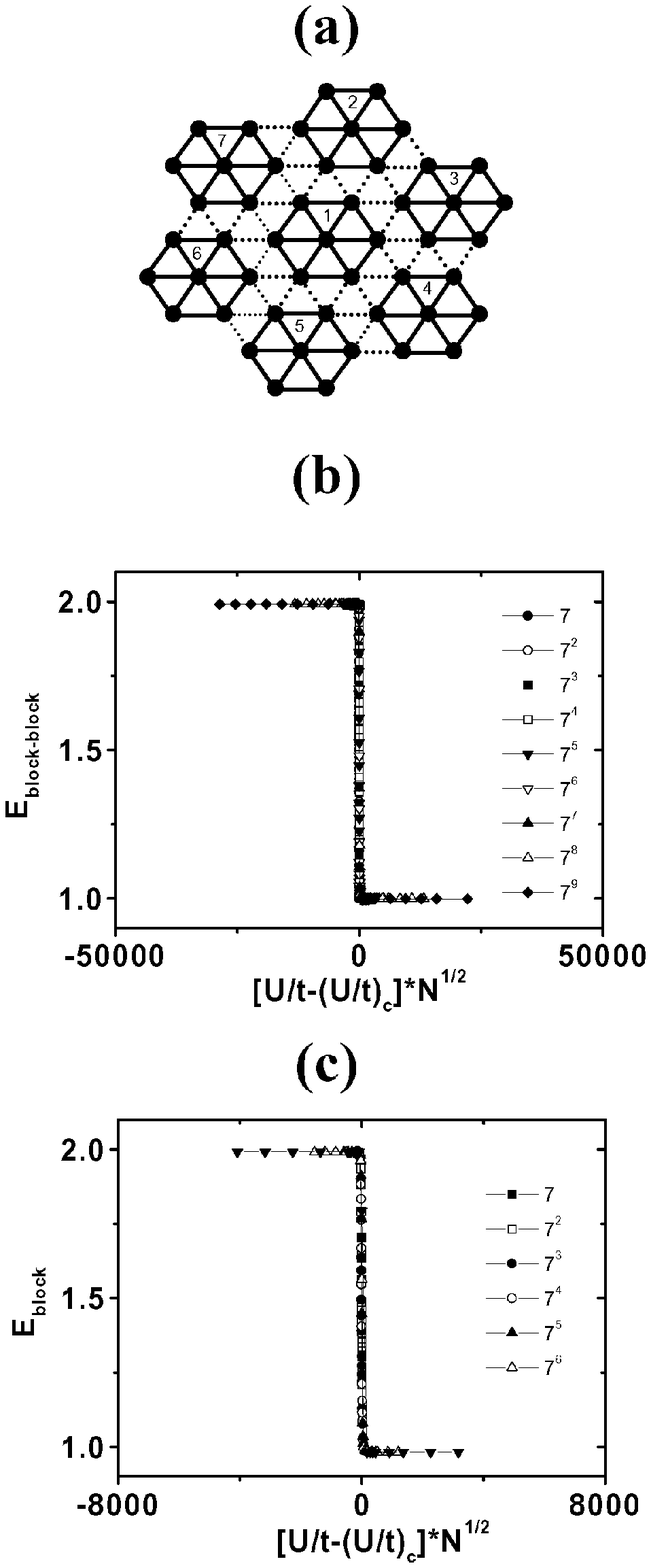}
\caption{Figure 2 : (a) Schematic diagram displays the lattice configuration with central purple block and the surrounding green ones. (b) Scaling of block-block  for various system size and (c) Scaling of block entanglements with the block size.} 
\end{figure}


\begin{references}
\bibitem{Mattle} K. Mattle, H. Weinfurter, P. G.Kwiat and A. Zeilinger, Phys. Rev. Lett.76,4546 (1996).
\bibitem{bouwmeester} D. Bouwmeester, J. Pan, K. Mattle, M. Eibl, H. Weinfurter and A. Zeilinger, Nature 390,575 (1997).
\bibitem{jennewein} T. Jennewein, C. Simon, G. Weihs, H. Weinfurter and A. Zeilinger, Phys. Rev. Lett. 84, 4729 (2000).
\bibitem{vandersypen} L. M. K. Vandersypen, M. Steffen, G. Breyta, C. S. Yannoni, M. H. Sherwood and I. L. Chuang, Nature 414, 883 (2001).
\bibitem{Lewenstein} M. Lewenstein, D. Bruss; J. I. Cirac, B. Kraus, M. Ku, J. Samsonowicz, A. Sanpera and R. Tarrach, J. Mod. Opt. 47, 2841 (2000).
\bibitem{bruss} D. Bruss, J. Math. Phys. 43, 4237 (2002).
\bibitem{lamas} A. Lamas-Linares, J. C. Howell and D. Bouwmeester, Nature 412, 887 (2001).
\bibitem{eibl} M. Eibl, S. Gaertner, M. Bourennane, C. Kurtsiefer, M. Zukowski and H. Weinfurter, Phys. Rev. Lett. 90, 200403 (2003).
\bibitem{berkley} A. J. Berkley, H. Xu, R. C. Ramos, M. A. Gubrud, F. W. Strauch, P. R. Johnson, J. R. Anderson, A. J. Dragt, C. J. Lobb, F. C. Wellstood, Nature 300, 1548 (2003).
\bibitem{wootters} W. K. Wootters, Phys. Rev. Lett. 80, 2245 (1998). 
\bibitem{schliemann} J. Schliemann, J. Ignacio Cirac, M. Kus, M. Lewenstein and D. Loss, Phys. Rev. A64, 022303 (2001).
\bibitem{li} Y. S. Li, B. Zeng, X. S. Liu and G. L. Long, Phys. Rev. A 64, 054302 (2001).
\bibitem{ahronov} D. Aharonov, Phys. Rev. A 62, 062311 (2000).
\bibitem{zanardi1} P. Zanardi and X. Wang, J. Phys. A: Math. Gen. 35, 7947 (2002).
\bibitem{gittings} J. R. Gittings and A. J. Fisher, Phys. Rev. A 66, 032305 (2002).
\bibitem{zanardi} P. Zanardi, Phys. Rev. A 65, 042101 (2002).
\bibitem{vidal} G. Vidal, J. I. Latorre, E. Rico and A. Kitaev, e-print, quant-ph/0211074.
\bibitem{sachdev} S. Sachdev, \textit{Quantum Phase Transitions} (Cambridge University Press, Cambridge, 1999).
\bibitem{osborne1} T. J. Osborne and M. A. Nielsen, Phys. Rev. A 66, 032110 (2002); e-print, quant-ph/0109024.
\bibitem{kais} O. Osenda, Z. Huang and S. Kais, Phys. Rev. A 67, 062321 (2003). 
\bibitem{osterloh} A. Osterloh, L. Amico, G. Falci and R. Fazio, Nature 416, 608 (2002).
\bibitem{white} S. R. White, Phys. Rev. Lett. 69, 2863 (1992).
\bibitem{vidal1} G. Vidal, quant-ph/0301063, 2003; quant-ph/0310089, 2003. Latorre J. I., E. Rico, G. Vidal, quant-ph/0304098, 2003. 
\bibitem{loss1} D. Loss, D. P. DiVincenzo, Phys. Rev. A 57, 120 (1998).
\bibitem{losss2} G. Burkard, D. Loss, D. P. DiVincenzo, Phys. Rev. B 59, 2070 (1999).
\bibitem{sarma} X. Hu, S. Das Sarma, Phys. Rev. A 61, 062301-1 (2000).
\bibitem{realspace} {\it{Real-Space Renormalization}} (Topics in Current Physics), Editors: T. W. Burkhardt and J.
M. J. van Leeuwen, Springer-Verlag, 1982.
\bibitem{wang1} J. X. Wang, S. Kais and R. D. Levine, Int. J. Mol. Sci. 3, 4 (2002).
\bibitem{wang2} J. X. Wang, S. Kais, Phys. Rev. B 66, 081101(R) (2002).
\bibitem{wang3} J. X. Wang, S. Kais, F. Remacle and R. D. Levine, J. Chem. Phys. B 106, 12847(2002).
\bibitem{rafi} F. Remacle, R. D. Levine, CHEMPHYSCHEM 2, 20 (2001).
\end{references}
\end{document}